\documentclass[prd,byrevtex
,preprint
,nofootinbib,
,tightenlines
,showkeys
,showpacs]{revtex4}
\usepackage{amsmath,amsxtra}
\usepackage{bm}
\usepackage[dvipdfm]{graphicx}
\usepackage{float}
\newcommand{\bs}[1]{\boldsymbol{#1}}
\newcommand{\pa}{\partial}

\newcommand{\ep}{\epsilon}
\begin{document}
\title{Classical and Quantum Correlations of Scalar Field in the
  Inflationary Universe}
\author{Yasusada Nambu}
\email{nambu@gravity.phys.nagoya-u.ac.jp}
\author{Yuji Ohsumi}
\email{osumi@gravity.phys.nagoya-u.ac.jp}
\affiliation{Department of Physics, Graduate School of Science, Nagoya 
University, Chikusa, Nagoya 464-8602, Japan}

\begin{abstract}
  We investigate classical and quantum correlations of a quantum field
  in the inflationary universe using a particle detector model. By
  considering the entanglement and correlations between two comoving
  detectors interacting with a scalar field, we find that the
  entanglement between the detectors becomes zero after their physical
  separation exceeds the Hubble horizon. Furthermore, the quantum
  discord, which is defined as the quantum part of total correlation,
  approaches zero on the super horizon scale. These behaviors support
  the appearance of the classical nature of the quantum fluctuation generated
  during the inflationary era.
\end{abstract}
\maketitle

\section{Introduction}
According to the inflationary scenario of cosmology, all structure in
the Universe can be traced back to primordial quantum fluctuations
during an accelerated expanding phase of the very early
universe. Short wavelength quantum fluctuations generated during
inflation are considered to lose quantum nature when their wavelengths
exceed the Hubble horizon length. Then, the statistical property of
generated fluctuations can be represented by classical
distribution functions. This is the assumption of the quantum to
classical transition of quantum fluctuations generated by the
inflation. As the structure of the present Universe is classical
objects, we must explain or understand how this transition occurred
and how the quantum fluctuations changed to classical fluctuations.

  We have investigated this problem from the viewpoint of
  entanglement~\cite{NambuY:PRD78:2008,NambuY:PRD80:2009}. For quantum
  fluctuations behaving in the  classical way, quantum expectation values
  of any operators must be calculable using appropriate classical
  distribution functions. Restricting to two point correlation
  functions of quantum operators, this condition is equivalent to the
  separability of a bipartite state. For a separable state, the
  entanglement is zero and there exists a positive normalizable
  P function~\cite{GardinerCW:S:2004} and this function can play the role of
  classical distribution function. In our previous study, we defined
  two spatially separated regions in the inflationary universe and
  investigated the bipartite entanglement between these regions. We
  found that the entanglement between these two regions becomes zero
  after their physical separation exceeds the Hubble horizon. This
  behavior of the bipartite entanglement confirms our expectation that
  the long wavelength quantum fluctuations during inflation behave as
  classical fluctuations and can become seed fluctuations for the
  structure formation in the Universe.  Our previous analysis
  concerning the entanglement of quantum fluctuations in the
  inflationary universe relies on the separability criterion for
  continuous bipartite
  systems~\cite{SimonR:PRL84:2000,DuanL:PRL84:2000} of which dynamical
  variables are continuous. The applicability of this criterion is
  limited to systems with Gaussian states: the wave function or the
  density matrix of the system is represented in a form of Gaussian
  function. Thus, we cannot say anything about the entanglement for
  the system with non-Gaussian states such as excited states and
  thermal states. Furthermore, from a viewpoint of observation or
  measurement, information on quantum fluctuations can be extracted
  via interaction between quantum fields and measurement
  devices. Hence, it is more natural to consider a setup in which the
  entanglement of quantum fields is probed using detectors.

  Following this direction, we consider particle
  detectors~\cite{UnruhWG:PRD14:1976,BirrellND:CUP:1982} with two
  internal energy levels interacting with a scalar field in this
  paper. By preparing two spatially separated equivalent detectors
  interacting with the scalar field, we can extract the information on
  entanglement of the scalar field by evaluating the entanglement
  between these two detectors. As a pair of such detectors is a
  two-qubit system, we have the necessity and sufficient condition for
  entanglement of this
  system~\cite{PeresA:PRL77:1996,HorodeckiM:PLA223:18}.  Using this
  setup, B. Reznik~\textit{el
    al.}~\cite{ReznikB:FP33:2003,ReznikB:PRA71:2005} studied the
  entanglement of the Minkowski vacuum. They showed that an initially
  nonentangled pair of detectors evolved to an entangled state
  through interaction with the scalar field. As the entanglement
  cannot be created by local operations, this implies that the
  entanglement of the quantum field is transferred to a pair of
  detectors. M.~Cliche and A.~Kempf~\cite{ClicheM:PRA81:2010}
  constructed the information-theoretic quantum channel using this
  setup and evaluated the classical and quantum channel capacities as
  a function of the spacetime separation.  G.~V.~Steeg and
  N.~C.~Menicucci~\cite{SteegGV:PRD79:2009} investigated the
  entanglement between detectors in de Sitter spacetime and they
  concluded that the conformal vacuum state of the massless scalar
  field can be discriminated from the thermal state using the
  measurement of entanglement.

  In this paper, we investigate the entanglement structure of the
  quantum field in the expanding universe using the particle detector
  model. We also consider correlations between detectors and explore
  the relation between classical and quantum parts of correlations.
  This paper is organized as follows. In Sec.~II, we present our setup
  of a detector system. Then, in Sec.~III, we review entanglement
  measure (negativity) and classical and quantum correlations for a
  two-qubit system. In Sec.~IV, we calculate entanglement and
  correlations for quantum fields in de Sitter spacetime and discuss
  how the classical nature of quantum fluctuations appears. Section~V is
  devoted to summary. We use units in which $c=\hbar=G=1$ throughout
  the paper.

\section{Two detectors system}
We consider a system with two equivalent detectors interacting with
the massless scalar field in an expanding
universe~\cite{ReznikB:PRA71:2005,SteegGV:PRD79:2009}. The detectors
have two energy level states $|\uparrow\rangle, |\downarrow\rangle$ and
their energy difference is given by $\Omega$. The interaction
Hamiltonian is assumed to be
\begin{equation}
  V=g(t)(\sigma_A^{+}+\sigma_A^{-})\phi(\bs{x}_A(t))
+g(t)(\sigma_B^{+}+\sigma_B^{-})\phi(\bs{x}_B(t))
\end{equation}
where $\sigma^{+},\sigma^{-}$ are raising and lowering operators for
the detector's state:
$$
 \sigma^{+}=|\uparrow\rangle\langle\downarrow|,\quad
 \sigma^{-}=|\downarrow\rangle\langle\uparrow|.
$$
Two detectors are placed at $\bs{x}_{A,B}(t)$ and
$(t,\bs{x}_{A,B}(t))$ represent their world lines. We assume that
the detectors are comoving with respect to cosmic expansion.  Strength of
the coupling is controlled in accord with  the following Gaussian window function
\begin{equation}
 g(t)=g_0\exp\left(-\frac{(t-t_0)^2}{2\sigma^2}\right).
\end{equation}
This window function approximates the detector being ``on'' when
$|t-t_0|\lesssim\sigma$ and ``off'' the rest of time. We assume that
the detectors are both down state initially $(t\rightarrow-\infty)$ and the
scalar field is vacuum state $|0\rangle$. Thus, the initial state of
the total system is
$|\Psi_0\rangle=|\downarrow\downarrow\rangle|0\rangle$. Then, in the
interaction representation, the final state $(t\rightarrow+\infty)$ of
the total system after interaction becomes
\begin{align}
  |\tilde\Psi\rangle&=\left[1-i\int_{-\infty}^{\infty} dt_1\tilde
    V_1-\frac{1}{2}\int_{-\infty}^{\infty}
    dt_1dt_2\mathrm{T}[\tilde V_1\tilde
    V_2]+\cdots\right]|\tilde\Psi_0\rangle \notag\\
  &=\left(1-\frac{1}{2}\mathrm{T}\left[\Phi_{A}^{-}\Phi_{A}^{+}+\Phi_{B}^{-}\Phi_{B}^{+}
\right]\right)|\downarrow\downarrow\rangle|0\rangle-i\Phi_{A}^{+}|\uparrow\downarrow
\rangle|0\rangle-i\Phi_B^{+}|\downarrow\uparrow\rangle|0\rangle\\
  &\qquad\qquad
  -\frac{1}{2}\mathrm{T}\left[\Phi_{A}^{+}\Phi_{B}^{+}+\Phi_{B}^{+}\Phi_{A}^{+}
    \right]|\uparrow\uparrow\rangle|0\rangle+O(g^3) \notag
\end{align}
where $\mathrm{T}$ is time ordering and symbols with a tilde denote quantities in the
interaction representation. We defined a field operator
$$
 \Phi_{A,B}^{\pm}=\int_{-\infty}^{\infty}dt_1 g(t_1)e^{\pm i\Omega
   t_1}\phi(t_1,\bs{x}_{A,B}(t_1)).
$$
As we are interested in the detectors' state, by tracing out the degrees
of freedom of the scalar field, the state for the detectors' degrees
of freedom becomes
\begin{equation}
  \label{eq:state2}
  \rho^{AB}=\begin{pmatrix}
    X_4 & 0 & 0 & X \\
    0 & E & E_{AB} & 0 \\
    0 & E_{AB} & E & 0 \\
    X^* & 0 & 0 & 1-2E-X_4
    \end{pmatrix}
\end{equation}
where we use the basis
$\{\uparrow\uparrow,\uparrow\downarrow,\downarrow\uparrow,
\downarrow\downarrow\}$ for this matrix representation of the state.
The matrix elements of the state \eqref{eq:state2} are
\begin{align*}
  &X=-2\int_{-\infty}^{\infty}dt_1\int_{-\infty}^{t_1} dt_2
  g_1g_2e^{i\Omega(t_1+t_2)}\langle\phi(t_1,\bs{x}_A)\phi(t_2,\bs{x}_B)\rangle, \\
  &E_{AB}=\int_{-\infty}^{\infty}dt_1\int_{-\infty}^{\infty} dt_2
  g_1g_2e^{-i\Omega(t_1-t_2)}\langle\phi(t_1,\bs{x}_A)\phi(t_2,\bs{x}_B)\rangle,\quad
  E=E_{AB}(r=0),\\
  &X_4=4\int_{-\infty}^{\infty}dt_3\int_{-\infty}^{t_3}dt_4g_3g_4\\
  &\quad\times\int_{-\infty}^{\infty}dt_1
  \int_{-\infty}^{t_1}dt_2g_1g_2e^{-i\Omega(t_3+t_4)}e^{i\Omega(t_1+t_2)}
  \langle\phi(t_4,\bs{x}_B)\phi(t_3,\bs{x}_A)\phi(t_1,\bs{x}_A)\phi(t_2,\bs{x}_B)\rangle.
\end{align*}
In $X_4$, the four point function can be written using two point
functions
\begin{align*}
  \langle\phi(t_4,\bs{x}_B)\phi(t_3,\bs{x}_A)\phi(t_1,\bs{x}_A)\phi(t_2,\bs{x}_B)\rangle
  &=\langle\phi(t_4,\bs{x}_B)\phi(t_1,\bs{x}_A)\rangle
  \langle\phi(t_3,\bs{x}_A)\phi(t_2,\bs{x}_B)\rangle\\
  &+\langle\phi(t_4,\bs{x}_B)\phi(t_3,\bs{x}_A)\rangle
  \langle\phi(t_1,\bs{x}_A)\phi(t_2,\bs{x}_B)\rangle\\
&+\langle\phi(t_3,\bs{x}_A)\phi(t_1,\bs{x}_A)\rangle
  \langle\phi(t_4,\bs{x}_B)\phi(t_2,\bs{x}_B)\rangle.
\end{align*}
$E, E_{AB}$ are $O(g^2)$ and $X_4$ is $O(g^4)$ quantities. For the
purpose of obtaining entanglement only, it is not necessary to
evaluate $X_4$. We need $X_4$ to obtain quantum mutual information
in which quantum part of correlations are encoded.  By changing
integration variables to
$$
 x_1=\frac{t_1+t_2}{2},\quad y_1=\frac{t_1-t_2}{2},\quad
 x_2=\frac{t_3+t_4}{2},\quad y_2=\frac{t_3-t_4}{2},
$$
we have the following expressions for the matrix elements:
\begin{align}
  &X=-4g_0^2e^{2i\Omega t_0}\int_{-\infty}^{\infty}dx
  e^{-x^2/\sigma^2+2i\Omega x}\int_0^\infty dy
  e^{-y^2/\sigma^2}D^{+}(x+t_0,y,r), \notag \\
  &E_{AB}=2g_0^2\int_{-\infty}^{\infty}dx
  e^{-x^2/\sigma^2}\int_{-\infty}^\infty dy
  e^{-y^2/\sigma^2-2i\Omega y}D^{+}(x+t_0,y,r),\notag \\
 &X_4=16g_0^4\int_{-\infty}^{\infty}dx_1e^{-x_1^2/\sigma^2+2i\Omega
   x_1}\int_0^\infty dy_1e^{-y_1^2/\sigma^2}
\int_{-\infty}^{\infty}dx_2e^{-x_2^2/\sigma^2-2i\Omega
   x_2}\int_0^\infty dy_2e^{-y_2^2/\sigma^2}  \label{eq:XE}\\
&\quad
\times
\Bigl[D^{+}\left(\frac{x_1+y_1+x_2-y_2}{2}+t_0,\frac{x_2-y_2-x_1-y_1}{2},r\right)
 \notag \\
 &\qquad\qquad\qquad\qquad 
 \times D^{+}\left(\frac{x_1-y_1+x_2+y_2}{2}+t_0,\frac{x_2+y_2-x_1+y_1}{2},r\right)
\notag \\
&\quad\quad +D^{+}(x_1+t_0,y_1,r)
D^{+}(x_2+t_0,y_2,r) \notag\\
&\quad\quad
+D^{+}\left(\frac{x_1+y_1+x_2+y_2}{2}+t_0,\frac{x_2+y_2-x_1-y_1}{2},0\right)
 \notag \\
&\qquad\qquad\qquad\qquad \times
D^{+}\left(\frac{x_1-y_1+x_2-y_2}{2}+t_0,\frac{x_2-y_2-x_1+y_1}{2},0\right)\Bigr] \notag
\end{align}
where we have introduced the Wightman function for the scalar field
$$
 D^{+}(x_1,y_1,r)\equiv\langle\phi(t_1,\bs{x}_A)\phi(t_2,\bs{x}_B)\rangle,\quad
 r=|\bs{x}_A-\bs{x}_B|.
$$
\section{Entanglement and correlations of detectors}
As the response of detectors due to interaction with the scalar field
is given by the state \eqref{eq:state2}, we can extract information on 
entanglement and correlations of the scalar field indirectly by
analyzing this state.

\subsection{Entanglement of detectors}
As a measure of the entanglement between two detectors, we consider
the negativity \cite{VidalG:PRA65:2002} defined via a partially
transposed operation to the density matrix \eqref{eq:state2} with
respect to detector B's degrees of freedom. The eigenvalues of the
partially transposed density matrix are
$$
  \lambda=E\pm|X|+O(g^4).
$$
The negativity is defined using the eigenvalues of the partially
transposed density matrix
\begin{equation}
  \mathcal{N}=\sum_{\lambda_i<0}|\lambda_i|=\mathrm{max}\Bigl[0, |X|-E\Bigr].
\end{equation}
From here on, we designate the following quantity as the negativity
\begin{equation}
\label{eq:negativity}
 \mathcal{N}=|X|-E.
\end{equation}
The negativity gives the necessity and the sufficient condition of the
entanglement for two-qubit
systems~\cite{PeresA:PRL77:1996,HorodeckiM:PLA223:18}. Thus two
detectors are entangled when $\mathcal{N}>0$ and separable when
$\mathcal{N}<0$. For separable initial states of detectors
$\mathcal{N}<0$, $\mathcal{N}>0$ after interaction with the scalar
field implies the scalar field is entangled because entanglement
cannot be generated by local operations.

\subsection{Correlations of detectors}

Using Bloch representation, the state \eqref{eq:state2} can be written
as follows
\begin{align}
 &\rho^{AB}=\frac{1}{4}\Biggl(I\otimes I+\bs{a}\cdot\bs{\sigma}\otimes
 I+I\otimes\bs{b}\cdot\bs{\sigma}+\sum_{\ell,m=1}^3c_{\ell
   m}\sigma_{\ell}\otimes\sigma_m\Biggr), \label{eq:state}\\
 &\bs{a}=\bs{b}=(0,0,-1+2E+2X_4),\qquad
 c_{\ell m}=\begin{pmatrix}
   2(E_{AB}+X_R) & -2X_I & 0 \\
   -2X_I & 2(E_{AB}-X_R) & 0 \\
   0 & 0 & 1-4E
   \end{pmatrix} \notag
\end{align}
where $I$ is the identity operator, $\{\sigma_1,\sigma_2,\sigma_3\}$
are the Pauli spin matrices, $X_R=\mathrm{Re}(X)$ and
$X_I=\mathrm{Im}(X)$.  To quantify quantumness of quantum fluctuations
in de Sitter spacetime, we want to consider the  ``classical'' and ``quantum''
part of the correlation between two detectors. The ``classical'' part of
correlation is defined through a local measurement on each
detector. By a measurement here we mean the von Neuman type; complete
measurement consisting of orthogonal one-dimensional projectors.

\subsubsection{Classical mutual information}
To define the classical part of the correlation between two detectors (two
qubits), we perform a local projective measurement of detector
states. Of course, it is not possible to perform measurement of the
scalar field in the inflationary era directly. We consider the
following measurement procedure as a gedanken experiment to explore
the nature of quantum fluctuation. The measurement operators for each
detector are
\begin{equation}
\label{eq:measurement}
 M_{\pm}^A=\frac{I\pm\bs{n}_A\cdot\bs\sigma}{2},\qquad
 M_{\pm}^B=\frac{I\pm\bs{n}_B\cdot\bs\sigma}{2},\qquad|\bs{n}_A|=|\bs{n}_B|=1
\end{equation}
where $\pm$ denotes output of the measurement. $\bs{n}_A, \bs{n}_B$
represent the internal direction of measurement.  The joint probability
$p_{jk}$ attaining measurement result $j$ for detector A and $k$ for
detector B ($j,k=\pm 1$) is obtained as
\begin{align}
  p_{jk}&=\mathrm{tr}(M_j^A\otimes M_k^B\,\rho^{AB}) \notag \\
  &=\frac{1}{4}\left[
    (1+(j)a_zn^A_z)(1+(k)b_zn^B_z)+(jk)\sum_{\ell m}(c_{\ell
      m}-a_\ell b_m)n^A_\ell n^B_m\right], \label{eq:pij}\\
  &\quad c_{\ell m}-a_\ell b_m=\begin{pmatrix}
    2(E_{AB}+X_R) & -2X_I & 0 \\
   -2X_I & 2(E_{AB}-X_R) & 0 \\
   0 & 0 & -4E^2+4X_4
   \end{pmatrix}. \notag
\end{align}
The probability $p_j$ attaining a result $j$ for detector A and $p_k$
attaining a result $k$ for detector B are
$$
  p_j=\sum_k p_{jk},\qquad p_k=\sum_j p_{jk}. 
$$
Using the joint probability \eqref{eq:pij} obtained by the measurement, the classical
mutual information $I_C$ is defined by~\cite{NielsenMA:CUP:2000}
$$
 I_C(p)=H(p_j)+H(p_k)-H(p_{jk}),\quad H(p_j)=-\sum_j p_j\log_2
 p_j,\quad
 H(p_{jk})=-\sum_{jk}p_{jk}\log_2p_{jk}
$$
where $H(p)$ is a Shannon entropy for a probability distribution $p$.
The explicit form of $I_C(p)$ using $p_{jk}$ is
\begin{equation}
 I_C(p)=\sum_{jk}p_{jk}\log_2\left(\frac{p_{jk}}{p_jp_k}\right).
\end{equation}
The measure of classical correlation is defined via the maximization
done over all possible projective measurements:\footnote{This
  definition of the classical correlation is based on two-sided
  measurements of correlations. We can also define the classical
  correlation using one-sided measurements of  correlations.}
\begin{equation}
 \mathcal{C}(p)=\sup_{\{\bs{n}_A,\bs{n}_B\}}I_C(p).
\end{equation}

We evaluate the classical mutual information of $p_{jk}$ separately
for the cases where the vectors $\bs{n}_{A,B}$ are parallel to the
$z$ axes or not. For the case where $\bs{n}_A$ and $\bs{n}_B$ are both
parallel to the $z$ axes, we have
\begin{equation}
 I_C(p)=\frac{1}{\ln2}\Bigl(E^2-X_4+X_4\ln\left(\frac{X_4}{E^2}\right)\Bigr)+O(g^6).
\end{equation}
For the case where one of $\bs{n}_A$ or $\bs{n}_B$ is parallel to the
$z$ axes,
\begin{equation}
 I_C(p)=O(g^6).
\end{equation}
For the case where both $\bs{n}_A$ and $\bs{n}_B$ are not parallel to
the $z$ axes,
\begin{align}
 &I_C(p)=\frac{1}{2\ln 2}\frac{\left(\sum_{\ell m=x,y}{\tilde c}_{\ell
       m}(n_A)_{\ell}(n_B)_m\right)^2}{(1-(n_{Az})^2)
(1-(n_{Bz})^2)}+O(g^6),\\
&\qquad\qquad
{\tilde{c}}_{\ell m}=
\begin{pmatrix}
   2(E_{AB}+X_R) & -2X_I  \\
   -2X_I & 2(E_{AB}-X_R)  
   \end{pmatrix}. \notag
\end{align}
By introducing new direction vectors
$$
\tilde{\bs{n}}_A=\frac{1}{\sqrt{(n_{Ax})^2+(n_{Ay})^2}}\begin{pmatrix}
  n_{Ax}\\  n_{Ay}\end{pmatrix},\qquad
\tilde{\bs{n}}_B=\frac{1}{\sqrt{(n_{Bx})^2+(n_{By})^2}}\begin{pmatrix}
  n_{Bx}\\  n_{By}\end{pmatrix},
$$
we obtain
\begin{equation}
  \label{eq:IC}
 I_C(p)=\frac{1}{2\ln 2}\left(\sum_{\ell m}{\tilde c}_{\ell
     m}(\tilde n_A)_{\ell}(\tilde n_B)_m\right)^2.
\end{equation}
As the eigenvalues of the matrix ${\tilde c}$ is $2(E_{AB}\pm
|X|)$, the maximum of classical mutual information for this case is given by
\begin{equation}
 I_{C,\text{max}}(p)=\frac{2}{\ln 2}\Bigl(|E_{AB}|+|X|\Bigr)^2.
\end{equation}
Thus, the classical correlation of the state \eqref{eq:state} obtained
by the local projective measurement is
\begin{equation}
  \label{eq:cc}
  \mathcal{C}=\mathrm{max}\left[\frac{1}{\ln2}\Bigl(E^2-X_4+X_4\ln\left(\frac{X_4}{E^2}
\right)\Bigr),~\frac{2}{\ln 2}\Bigl(|E_{AB}|+|X|\Bigr)^2\right].
\end{equation}

We comment on the relation between the classical mutual information
and a correlation function of qubit variables. By the local projective
measurement of the state \eqref{eq:state}, we obtain the following
expectation values for qubit variables:
\begin{align*}
  &\langle\bs{n}_A\cdot\bs{\sigma}\rangle=\bs{a}\cdot\bs{n}_A,\qquad
  \langle\bs{n}_B\cdot\bs{\sigma}\rangle=\bs{b}\cdot\bs{n}_B,\\
  &\langle\bs{n}_A\cdot\bs{\sigma}\otimes\bs{n}_B\cdot\bs{\sigma}\rangle
  =\sum_{\ell,m}c_{\ell m}(n_A)_\ell(n_B)_m.
\end{align*}
By introducing a fluctuation part of  qubit variables as
$
 \Delta\sigma_{\bs{n}}=\bs{n}\cdot\bs{\sigma}-\langle\bs{n}\cdot\bs{\sigma}\rangle
$, 
the correlation function for fluctuations of qubit variables is
\begin{equation}
 \langle\Delta\sigma_{\bs{n}_A}\Delta\sigma_{\bs{n}_B}\rangle=\sum_{\ell,m}(c_{\ell
   m}-a_{\ell}b_m)(n_A)_\ell(n_B)_m=\sum_{\ell,m}\tilde c_{\ell
   m}(\tilde n_A)_\ell(\tilde n_B)_m
\end{equation}
where we have assumed $\bs{n}_A, \bs{n}_B$ are not parallel to $z$
axis at the last equality.  Therefore, the classical mutual
information \eqref{eq:IC} for joint probability $p_{jk}$ corresponds
to the square of the correlation function of the fluctuation part of
the qubit variables.

\subsubsection{Quantum mutual information and quantum discord}
The quantum mutual information $I_Q$ of the bipartite state
$\rho^{AB}$ is defined independently of measurement procedure:
\begin{equation}
 I_Q(\rho^{AB})=S(\rho^A)+S(\rho^B)-S(\rho^{AB}),\qquad
 S(\rho)=-\mathrm{tr}(\rho\log_2\rho)
\end{equation}
where $S(\rho)$ is the von Newmann entropy for the state $\rho$. This
quantity represents the total correlations of the bipartite system
including both the quantum and classical parts of correlations.

We evaluate the quantum mutual information for the state
\eqref{eq:state}. The eigenvalues of the state \eqref{eq:state} are
$1-2E+|X|^2-X_4, E\pm E_{AB}, X_4-|X|^2$. The reduced density matrix
for the subsystem A is
$$
 \rho^A=\mathrm{tr}_B\rho^{AB}=\begin{pmatrix} E+X_4 & 0 \\ 0 & 1-E-X_4 \end{pmatrix}.
$$
Thus, the quantum mutual information for the two detector system is
given by
\begin{align}
  I_Q(\rho)&=
   \frac{1}{\ln 2}\,\Bigl[-2E\ln E+(E+E_{AB})\ln(E+E_{AB})
  +(E-E_{AB})\ln(E-E_{AB})\Bigr] \label{eq:IQ}\\
  &\qquad +\frac{1}{\ln 2}\Bigl[E^2-X_4+|X|^2-2X_4\ln
  E+(X_4-|X|^2)\ln(X_4-|X|^2)\Bigr] \notag
  +O(g^6).
\end{align}

The quantum discord~\cite{OllivierH:PRL88:2002, VedralV:PRL90:2003,LuoS:JSP136:2009} is
introduced as the difference between the quantum mutual information
and the classical mutual information
\begin{equation}
 \mathcal{Q}(\rho)=I_Q(\rho)-\mathcal{C}(p).
\end{equation}
For arbitrary local projective measurements, it can be shown that
$\mathcal{Q}\geq 0$. Thus, the quantum mutual information can be decomposed
into a positive classical mutual information and a positive quantum
discord. The necessity and sufficient condition for zero quantum
discord is expressed as the state has the following
form~\cite{XuJW:X11013408:2011}:
\begin{equation}
 \rho^{AB}=\sum_{j,k}p_{jk}M^A_jM^B_k. \label{eq:zero-discord}
\end{equation}
For this state, measurement \eqref{eq:measurement} does not alter the
form of state because $M_j^A, M_k^B$ are projection operators. In this
sense, the state with zero quantum discord can be termed a classical
state and  the quantum discord represents the quantum part of the total
correlations. For pure state, we have $\mathcal{Q}=\mathcal{C}$ and
the state with zero quantum discord has no classical correlations.

\subsubsection{Bell-CHSH inequality}
Related to classical correlations obtainable via local measurements,
we consider the question whether correlations derived under the state
\eqref{eq:state} admit a local hidden-variable (LHV) description;
measured correlation functions can be mimicked by classical distribution
functions.  Let us consider the following operator (Bell operator):
\begin{equation}
  \mathcal{B}_{\text{CHSH}}=\bs{a}\cdot\bs{\sigma}\otimes(\bs{b}+\bs{b}')\cdot\bs{\sigma}
  +\bs{a}'\cdot\bs{\sigma}\otimes(\bs{b}-\bs{b}')\cdot\bs{\sigma}
\end{equation}
where $\bs{a},\bs{a}',\bs{b},\bs{b}'$ are real unit vectors. Then,
the Bell-Clauser-Horne-Shimony-Holt(CHSH) inequality \cite{ClauserJF:PRL23:1969} is
\begin{equation}
  |\langle\mathcal{B}_{\text{CHSH}}\rangle|\leq 2.
\end{equation}
If the state admits a LHV description of
correlations, then  this inequality holds.  Violation of
this inequality means existence of nonlocality. The two qubit state
violates the Bell-CHSH inequality if and only if the following
condition is satisfied~\cite{HorodeckiR:PLA200:1995} :
\begin{equation}
M(\rho)>1,\quad
  M(\rho)\equiv\text{the sum of the two largest eigenvalues of the
    matrix $c^{\dag}c$}
\end{equation}
where the matrix $c_{jk}=\mathrm{tr}(\rho\sigma_j\otimes\sigma_k)$.
For the state \eqref{eq:state},
\begin{align*}
 &c_{jk}=\begin{pmatrix} c_{11} & c_{12} & 0 \\ c_{12} & c_{22} & 0 \\
   0 & 0 &  c_{33} \end{pmatrix}, \quad c_{11}=2(E_{AB}+X_R),\\
&
 c_{22}=2(E_{AB}-X_R),\quad c_{12}=-2X_I,\quad c_{33}=1-4E.
\end{align*}
As eigenvalues of $c_{jk}$ are $1-4E, 2(E_{AB}\pm |X|)$, the sum of
the square of the two largest eigenvalues of $c$ cannot exceed unity
and the Bell-CHSH inequality holds. However, this does not mean the
LHV description of correlations is possible; holding the Bell-CHSH
inequality is only a necessary condition for the LHV description and
does not guarantee existence of a LHV.  Actually, by passing each
detector through the filter
$$
 f_{A,B}=\begin{pmatrix} 1 & 0 \\ 0 & \eta \end{pmatrix},
$$
there is a possibility revealing hidden nonlocality of the state
\cite{GisinN:PLA210:1996}. After passing through the filter, the state
is transformed according to $\rho\rightarrow \rho'=(f_A\otimes
f_B)\rho(f_A\otimes f_B)$. Assuming that $\eta^2=O(g^2)$, the
transformed normalized state is
\begin{equation}
  \frac{\rho'}{\mathrm{tr}\rho'}=\frac{1}{X_4+2\eta^2 E+\eta^4}\begin{pmatrix}
 X_4 & 0 & 0 & \eta^2 X \\
 0 & \eta^2 E & \eta^2 E_{AB} & 0 \\
 0 & \eta^2 E_{AB} & \eta^2 E & 0 \\
 \eta^2 X^* & 0 & 0 & \eta^4 \end{pmatrix}
\end{equation}
and the matrix $c$ is transformed to
\begin{equation}
  c'=\frac{2\eta^2}{X_4+2\eta^2 E+\eta^4}
  \begin{pmatrix}
    E_{AB}+X_R & -X_I & 0 \\
    -X_I & E_{AB}-X_R & 0 \\
    0 & 0 & \dfrac{X_4-2\eta^2 E+\eta^4}{2\eta^2}
  \end{pmatrix}.
\end{equation}
Assuming that the state is entangled $\mathcal{N}=|X|-E>0$, eigenvalues of $c'$ are
$$
 \pm\frac{2\eta^2|X|}{X_4+\eta^4},\quad
 1-\frac{4E\eta^2}{X_4+\eta^4}
$$
and the condition for violation of the Bell-CHSH inequality $M(\rho')>1$ is
$$
 \eta^4-\frac{|X|^2}{2E}\eta^2+X_4<0.
$$
For existence of a real $\eta$ satisfying this inequality, we
need~\cite{ReznikB:FP33:2003}
\begin{equation}
  |X|^4>16X_4E^2.
  \label{eq:bell}
\end{equation}
This provides a sufficient condition for violation of the Bell-CHSH
inequality and existence of a hidden nonlocality.

\newpage
\section{Behavior of entanglement and correlations of scalar field}
The matrix elements $E_{AB}, X, X_4$ of \eqref{eq:XE} can be evaluated
using numerical integration after considering contributions of poles
in integrands by contour integration on a complex
plane~\cite{OhsumuY:2011}. However, numerical integrations of these
functions are not so easy especially evaluating the fourfold integral of
$X_4$. Hence, in this paper, we consider asymptotic estimation of
these functions to derive analytic approximate forms.  We assume
parameters $\Omega\sigma\gg1$ with other dimensionless combinations of
parameters such as $H\Omega\sigma^2$ contained in $D^{+}$ are kept
order unity. That is, we are considering the asymptotic behavior of
$E_{AB}, X, X_4$ in the range of parameters
$$
 \Omega\sigma\gg1,\qquad H\sigma\ll 1
$$
where $H$ is the Hubble parameter.  By rescaling the integration variables
$x$ and $y$,
\begin{align*}
  &X=-4g_0^2(\Omega\sigma^2)^2e^{2i\Omega
    t_0}\int_{-\infty}^{\infty}dx
  e^{-(\Omega\sigma)^2(x^2-2ix)}\int_0^{\infty}dy
  e^{-(\Omega\sigma)^2y^2}D^{+}(t_0+\Omega\sigma^2
  x,\Omega\sigma^2y,r),\\
  &E_{AB}=2g_0^2(\Omega\sigma^2)^2\int_{-\infty}^{\infty}dx
  e^{-(\Omega\sigma)^2x^2}\int_{-\infty}^{\infty}dy
  e^{-(\Omega\sigma)^2(y^2+2iy)}D^{+}(t_0+\Omega\sigma^2
  x,\Omega\sigma^2y,r).
\end{align*}
For $\Omega\sigma\gg 1$, $x$ and $y$ integrals in $X$ can be
evaluated approximately at the saddle point $x=i$ and $y=0$ of the
integrand. After performing Gaussian integrals about these saddle
points, we have
\begin{align*}
  X&\approx -4g_0^2(\Omega\sigma^2)^2e^{2i\Omega
    t_0}\int_{-\infty}^{\infty}dx
  e^{-(\Omega\sigma)^2(x^2+1)}\int_0^{\infty}dy
  e^{-(\Omega\sigma)^2y^2}D^{+}(t_0+i\Omega\sigma^2,0,r) \\
  &=-2\pi g_0^2\sigma^2 e^{-(\Omega\sigma)^2}e^{2i\Omega
    t_0}D^{+}(t_0+i\Omega\sigma^2, 0, r).
\end{align*}
In the same way, expression for $E_{AB}$ is
\begin{align*}
  E_{AB}&\approx 2g_0^2(\Omega\sigma^2)^2\int_{-\infty}^{\infty}dx
  e^{-(\Omega\sigma)^2x^2}
  \int_{-\infty}^{\infty}dy e^{-(\Omega\sigma)^2(y^2+1)}D^{+}(t_0,
  -i\Omega\sigma^2,r)\\
  &=2\pi g_0^2\sigma^2e^{-(\Omega\sigma)^2}D^{+}(t_0,-i\Omega\sigma^2,r).
\end{align*}
Thus, the asymptotic forms of $X, E_{AB}$ for $\Omega\sigma\gg 1$ are
obtained as
\begin{align}
  &X\approx -2\pi
  g_0^2\sigma^2e^{-(\Omega\sigma)^2}D^{+}(t_0+i\Omega\sigma^2, 0, r),
  \notag \\
  &E_{AB}\approx 2\pi
  g_0^2\sigma^2e^{-(\Omega\sigma)^2}D^{+}(t_0,-i\Omega\sigma^2,r), \label{eq:asymp}\\
  &X_4\approx 4\pi^2g_0^4\sigma^4e^{-2(\Omega\sigma)^2}\Bigl[
  D^{+}(t_0,-i\Omega\sigma^2,r)^2+\left|D^{+}(i\Omega\sigma^2+t_0,0,r)\right|^2
  +D^{+}(t_0,-i\Omega\sigma^2,0)^2\Bigr] \notag \\
  &\qquad=E^2+E_{AB}^2+|X|^2. \notag
\end{align}
The negativity is
\begin{equation}
  \mathcal{N}\approx 2\pi g_0^2\sigma^2
  e^{-(\Omega\sigma)^2}\Bigl[|D^{+}(t_0+i\Omega\sigma^2,0,
  r)|-D^{+}(t_0, -i\Omega\sigma^2,0)\Bigr].
\end{equation}
Using the asymptotic form of $X_4$, we can reduce the condition of
hidden nonlocality \eqref{eq:bell}.  For the entangled region
$|X|\gg E$, using the asymptotic form of $X_4\approx
E^2+|X|^2+E_{AB}^2\approx |X|^2$, the condition of hidden nonlocality
\eqref{eq:bell} reduces to
\begin{equation}
 |X|>4E
\end{equation}
and provides a stronger condition than that for entanglement $|X|>E$.

Now, let us we consider entanglement and correlations of the scalar
fields in  Minkowski and de Sitter spacetime.
\subsection{Minkowski vacuum}
\begin{figure}[H]
  \centering
  \includegraphics[width=0.5\linewidth,clip]{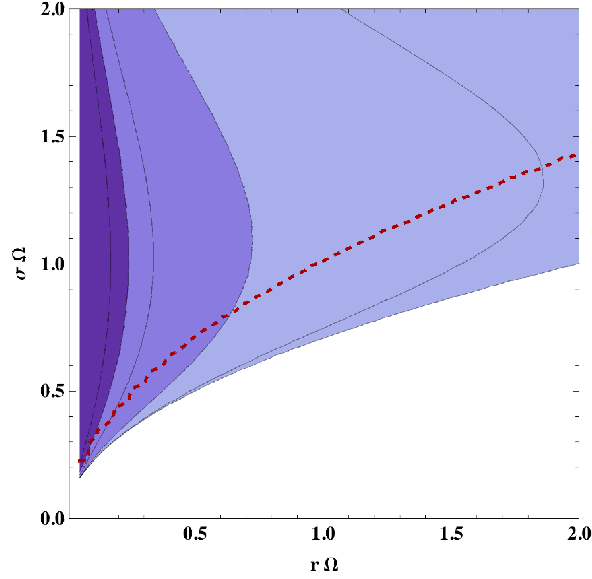}
  \caption{Contours of negativity for the Minkowski vacuum. Two
    detectors are entangled with parameters in the shaded
    region. Darker areas correspond to larger negativities. In the
    region above the dotted line, the Bell-CHSH inequality is violated
    after operation of local filtering.}
  \label{fig:minkow1}
\end{figure}

\begin{figure}[H]
  \centering
  \includegraphics[width=0.45\linewidth,clip]{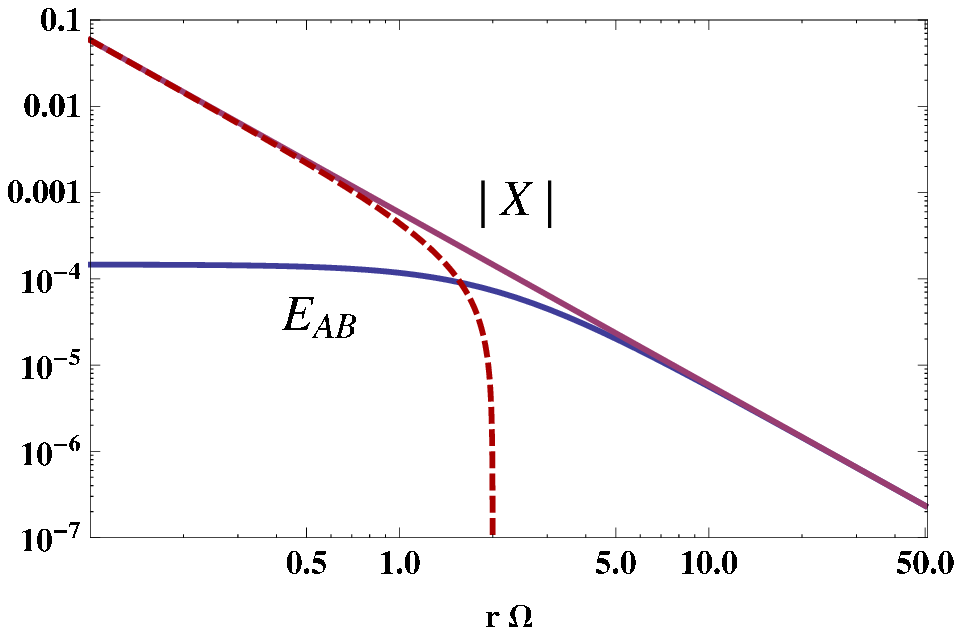}
  \includegraphics[width=0.45\linewidth,clip]{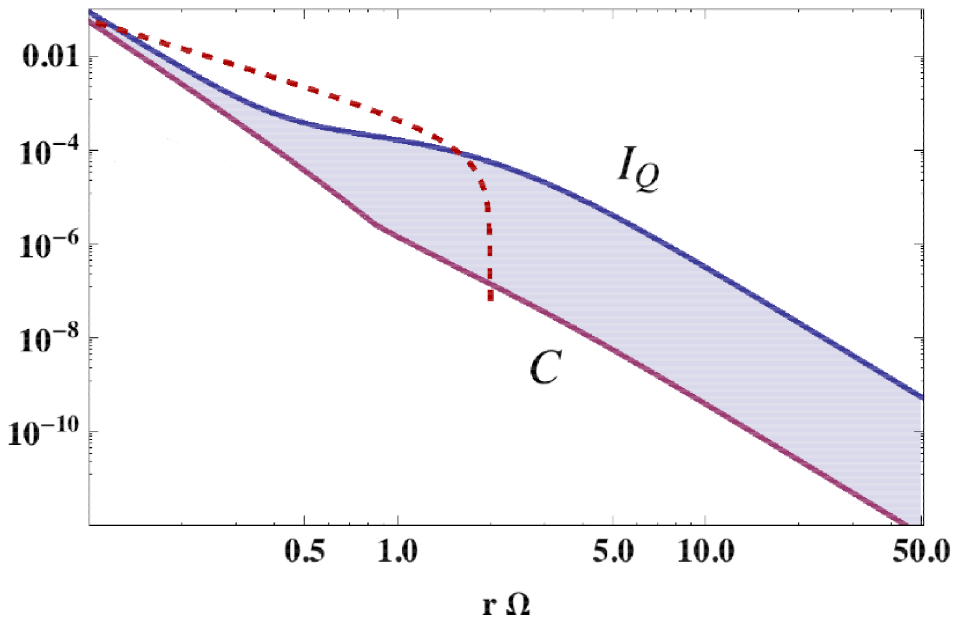}
  \caption{$r$ dependence of correlations
    ($\sigma\Omega=1,g_0=0.1$). The right panel shows $r$ dependence
    of $I_{Q}$ and $\mathcal{C}$. In both panels, dashed lines
    represent negativity.}
  \label{fig:minkow2}
\end{figure}

The Wightman function of the massless scalar with Minkowski vacuum
state is
\begin{equation}
 D^{+}=-\frac{1}{4\pi^2}\frac{1}{4(y-i\ep)^2-r^2}
\end{equation}
where a positive small constant $\ep>0$ is introduced to regularize
ultraviolet divergence in the Wightman function (see Appendix). Using \eqref{eq:asymp},
$$
X=-(2\pi
g_0^2)\frac{e^{-(\Omega\sigma)^2}}{4\pi^2}\left(\frac{\sigma}{r}\right)^2,\qquad
E_{AB}=(2\pi
g_0^2)\frac{e^{-(\Omega\sigma)^2}}{4\pi^2}\frac{1}{(r/\sigma)^2+4(\Omega\sigma)^2}.
$$
The negativity is
\begin{equation}
  \mathcal{N}=\frac{g_0^2}{8\pi}e^{-\Omega^2\sigma^2}\left[
    4\left(\frac{\sigma}{r}\right)^2-\frac{1}{(\Omega\sigma)^2}\right]
\end{equation}
and the $\mathcal{N}=0$ line is given by
\begin{equation}
  \sigma\Omega=\frac{r}{2\sigma}.
\end{equation}
For $r/\sigma<2\Omega\sigma$, the negativity is positive and detectors
are entangled (Fig.~\ref{fig:minkow1}). As the prepared initial state
of detectors is separable and entanglement cannot be created via local
operations, this entanglement is due to the scalar field and it is
swapped to a pair of detectors through interaction. For large
separation $r/\sigma>2\Omega\sigma$, detectors are separable. However,
this does not mean the scalar field is separable for this separation;
by increasing detector's parameter $\sigma$ with fixed $r$, it is
possible to make $\mathcal{N}>0$. For any separation $r$, we can
always choose an appropriate value of $\sigma$ that makes detectors
entangled. Thus, we can say that the Minkowski vacuum is always
entangled.
Now, let us consider behavior of correlations. The left panel in
Fig.~\ref{fig:minkow2} shows $r$ dependence of $|X|, E_{AB}$. They are
monotonically decreasing functions with respect to $r$. In the
entangled region $r<2\Omega\sigma^2$, $E_{AB}$ is nearly
constant $E_{AB}\approx E$ and in the separable region $r>2\Omega\sigma^2$,
$E_{AB}\approx |X|<E$.
\noindent
$r$ dependence of the quantum mutual information $I_Q$ and the
classical correlation $\mathcal{C}$ is shown in the right panel in
Fig.~\ref{fig:minkow2}. These correlations also decrease monotonically
with respect to $r$. In the entangled regions, the difference of these
two correlations (quantum discord) increases as $r$ increases until
two detectors become separable.  In the separable region, quantum discord
remains constant.
\begin{figure}[H]
  \centering
  \includegraphics[width=0.5\linewidth,clip]{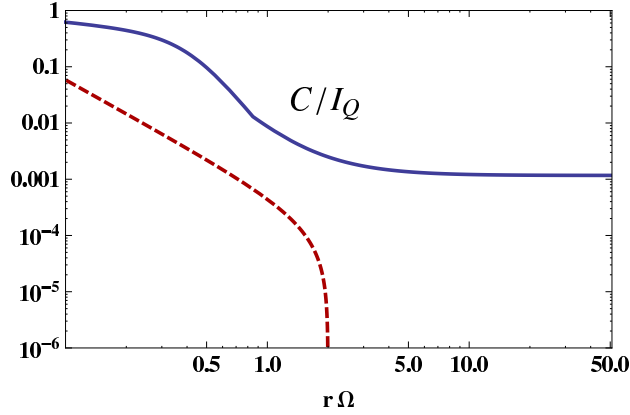}
  \caption{$r$ dependence of the ratio $\mathcal{C}/I_Q$
    ($\sigma\Omega=1,g_0=0.1$).  The dashed line represents
    negativity.}
  \label{fig:ratio-minkow}
\end{figure}
As shown in Fig.~\ref{fig:ratio-minkow}, the ratio of classical
correlation to total correlation $\mathcal{C}/I_Q$ is a  good
indicator of separability for this system. In the entangled region,
the ratio decreases monotonically and in the  separable region, the ratio
approaches a constant value. We can derive this behavior using the
following $r$ dependence of $X, E_{AB}$:
\begin{alignat}{3}
  &r\ll r_c:\qquad E_{AB}\approx E \ll |X| &\quad (\text{entangled
    region}), \label{eq:EX2} \\
  &r\gg r_c:\qquad |X|\approx E_{AB}\ll E &\quad (\text{separable
    region}), \notag
\end{alignat}
where $r_c=2\Omega\sigma^2$. For $r\ll r_c$,
\begin{equation}
  \mathcal{C}\approx\mathrm{max}\left\{\frac{2}{\ln
      2}|X|^2\ln\frac{|X|}{E}, \frac{2}{\ln 2}|X|^2 \right\}
,\qquad
  I_Q\approx -\frac{2}{\ln 2}|X|^2\ln E.
\end{equation}
In this region, as $|X|\gg E$, the ratio is
$$
 \frac{\mathcal{C}}{I_Q}\approx 1-\frac{\ln|X|}{\ln E}
$$
and this reproduces decreasing behavior of the ratio with respect to
$r$.  For $r\gg r_c$,
\begin{equation}
 \mathcal{C}\approx \frac{8}{\ln 2}E_{AB}^2,\qquad
 I_Q\approx\frac{2}{\ln 2}\frac{E_{AB}^2}{E}
\end{equation}
and the ratio $\mathcal{C}/I_Q$ approaches a constant value
independent of $r$ and its value is given by
\begin{equation}
\label{eq:ratio}
 \frac{\mathcal{C}}{I_Q}\approx 4E\sim g_0^2\frac{e^{-(\Omega\sigma)^2}}{(\Omega\sigma)^2}.
\end{equation}
The behavior of $r$ dependence of the ratio $\mathcal{C}/I_Q$ changes
at $|X|\approx E$ and this corresponds to the separable line
$\mathcal{N}=0$. The separability condition $\mathcal{N}<0$ comes from
the positive partially transposed criterion of the
state~\cite{VidalG:PRA65:2002}. Although the quantity
$\mathcal{C}/I_Q$ is introduced independently of the positive
partially transpoed criterion, the change of its $r$ dependence at
$\mathcal{N}\approx 0$ arises and it corresponds to the separability
condition.

\subsection{Scalar field in de Sitter spacetime}
In de Sitter spacetime with a spatially flat slice
$$
 ds^2=-dt^2+e^{2Ht}d\bs{x}^2,
$$
the Wightman function of a massless conformal scalar field with the
conformal vacuum state~\cite{BirrellND:CUP:1982} is
\begin{equation}
 D^{+}_{\text{conf}}=
 \frac{H^2}{16\pi^2}\left[-\sinh^2(H(y-i\ep))+e^{2Hx}(Hr/2)^2\right]^{-1}.
\end{equation}
The Wigthman function of a massless minimal scalar field with the
Bunch-Davis vacuum state is
\footnote{$\mathrm{Ei}(x)=-\int_x^{\infty}\frac{dk}{k}e^{-k}=\gamma+\ln
  x+\sum_{n=1}^{\infty}c_nx^n$.}
\begin{align}
  &D^{+}_{\text{min}}=D_{\text{conf}}^{+}+D_2, \\
  &D_2=-\frac{H^2}{8\pi^2}\left[\mathrm{Ei}\left[i(Hr-2e^{-Hx}\sinh(H(y-i\ep)))\right]
    +\mathrm{Ei}\left[i(-Hr-2e^{-Hx}\sinh(H(y-i\ep)))\right]\right].\notag
\end{align}
The massless minimal scalar field in de Sitter spacetime suffers from
infrared divergence. Hence we have introduced the infrared cutoff
$k_0=H$ in $D_2$ which corresponds to a maximal comoving size of
the inflating region; $r$ is the comoving distance between two spatially
separated  points and satisfies $r<H^{-1}$. We have assumed that
inflation starts at $t=0$ and the comoving distance $r$ must be smaller
than the size of the inflationary universe $H^{-1}$ at $t=0$.  The
asymptotic form of $X, E_{AB}$ are
\begin{alignat}{2}
 &X_{\text{conf}}=X_1,&\qquad &E_{AB}{}_{\text{conf}}=E_{AB}{}_1,\\
 &X_{\min}=X_1+X_2,&\qquad &E_{AB}{}_{\text{min}}=E_{AB}{}_1+E_{AB}{}_2 \notag
\end{alignat}
where
\begin{align}
  &X_1=-(2\pi g_0^2)
\frac{e^{-(\Omega\sigma)^2}}{4\pi^2}\frac{(H\sigma)^2}{(Hr_p)^2}e^{-2iH\Omega\sigma^2}
, \notag \\
  &E_{AB}{}_1=(2\pi g_0^2)
\frac{e^{-(\Omega\sigma)^2}}{4\pi^2}\frac{(H\sigma)^2}{4\sin^2(H\Omega\sigma^2)
  +(Hr_p)^2},\\
 &X_2=(2\pi g_0^2)\frac{e^{-(\Omega\sigma)^2}}{8\pi^2}(H\sigma)^2\left\{
   \mathrm{Ei}\left(-ie^{-Ht_0}Hr_p\right)+\mathrm{Ei}\left(ie^{-Ht_0}Hr_p\right)\right\},
 \notag \\
&E_{AB}{}_2=-(2\pi g_0^2)\frac{e^{-(\Omega\sigma)^2}}{8\pi^2}(H\sigma)^2\Bigl\{
   \mathrm{Ei}\left(-ie^{-Ht_0}Hr_p+2e^{-Ht_0}\sin(\Omega
     H\sigma^2)\right) \notag \\
&\qquad\qquad\qquad\qquad\qquad
+\mathrm{Ei}\left(ie^{-Ht_0}Hr_p+2e^{-Ht_0}\sin(\Omega
  H\sigma^2)\right)\Bigr\} \notag
\end{align}
where $Ht_0$ is the e-folding time at which detectors are switched on. The
physical distance between detectors at this instance is
$r_p=e^{Ht_0}r_0$ with $r_0<H^{-1}$. Unlike the Minkowski vacuum case,
physical distance between two detectors increases in accord with
cosmic expansion and entanglement between detectors changes in time.
\begin{figure}[H]
  \centering
 \includegraphics[width=0.45\linewidth,clip]{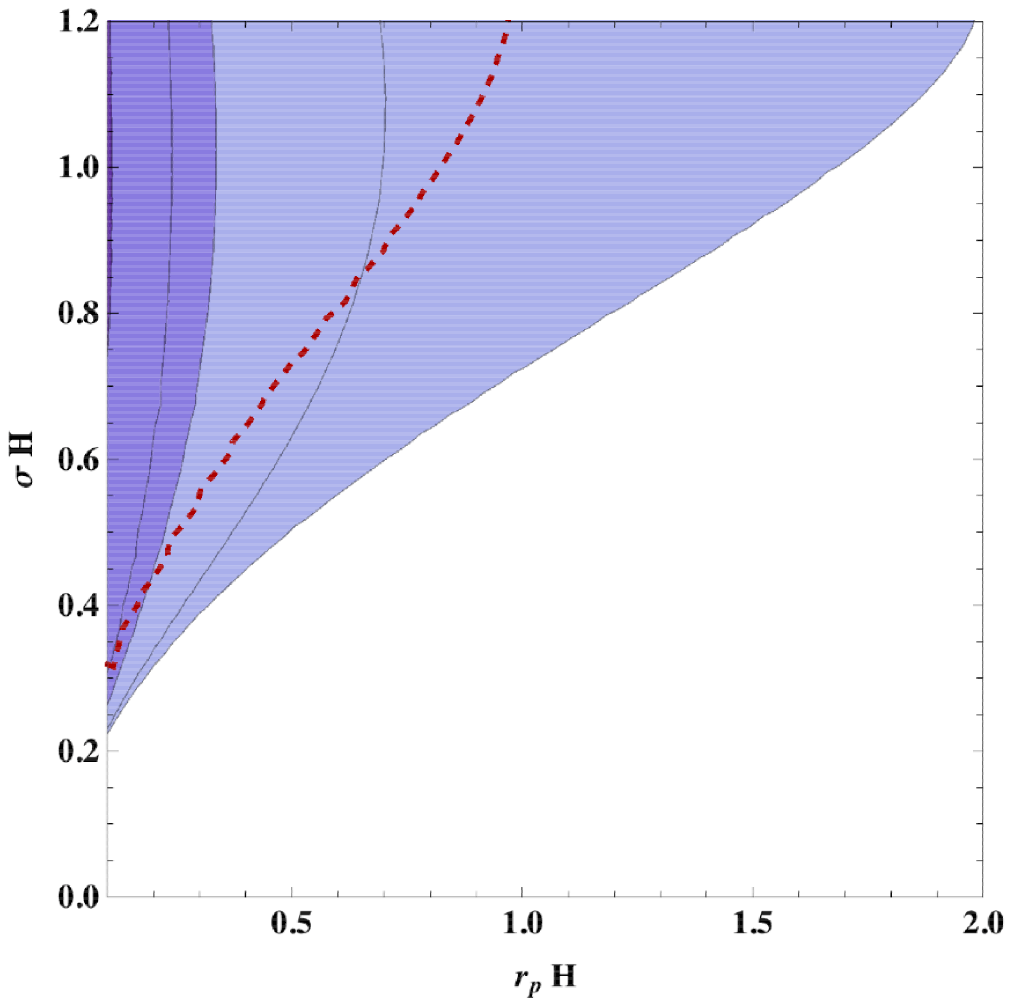}%
 \hspace{0.5cm}
  \includegraphics[width=0.45\linewidth,clip]{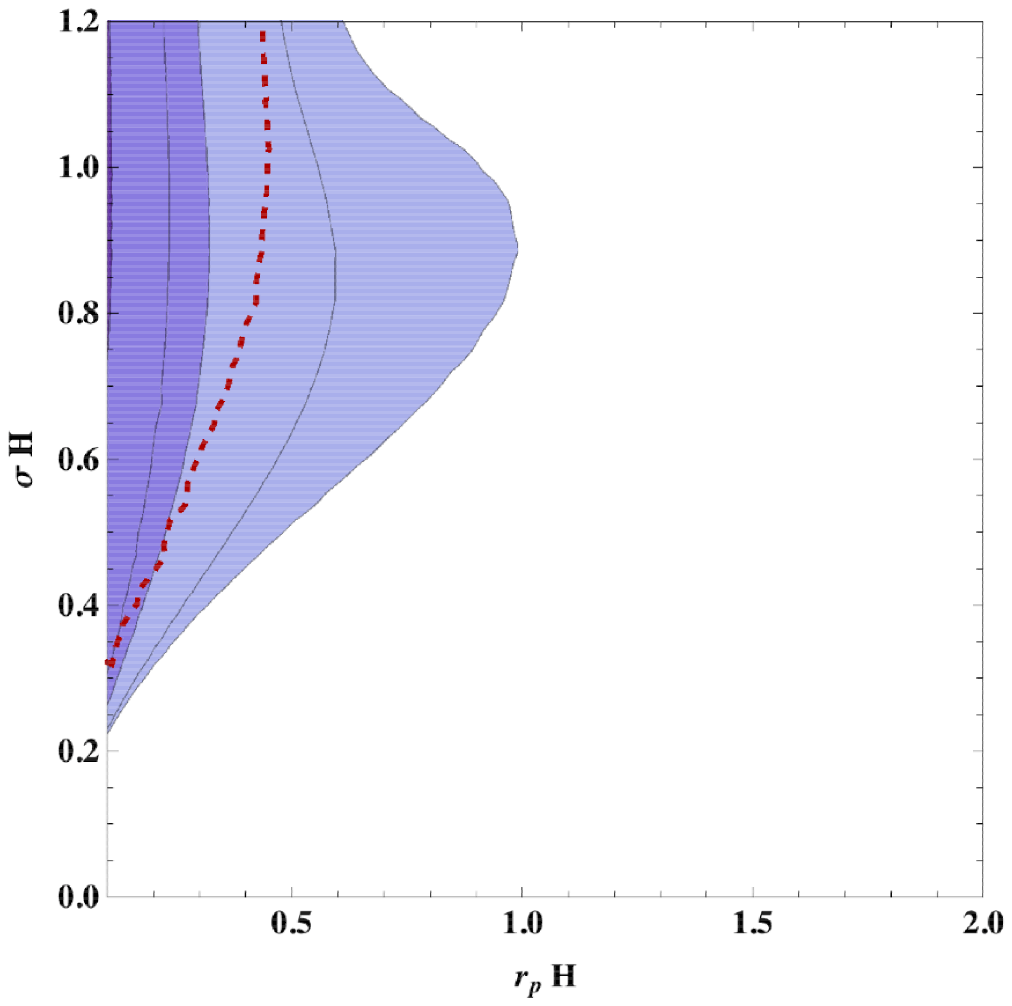}
  \caption{Contours of negativity for the conformal invariant scalar
    field (left panel) and the minimal scalar field (right
    panel) ($\Omega/H=1, r_0=0.1$). Two detectors are entangled with
    parameters in the shaded region. In the region above the dotted
    line, the Bell-CHSH inequality is violated after operation of
    local filtering.}
\end{figure}
For sufficiently large e-foldings $Ht_0\gg 1$,
\begin{align*}
  &X_{\text{min}}\approx X_2\approx
  -(2\pi g_0^2)\frac{e^{-(\Omega\sigma)^2}}{4\pi^2}(H\sigma)^2\left[-\ln(Hr_p)+Ht_0\right],\\
  &E_{AB}{}_{\text{min}}\approx E_{AB}{}_2\approx(2\pi g_0^2)
\frac{e^{-(\Omega\sigma)^2}}{4\pi^2}(H\sigma)^2\left[\frac{1}{4\sin^2(H\Omega\sigma^2)}-
    \ln\sqrt{(Hr_p)^2+4\sin^2(H\Omega\sigma^2)}+Ht_0\right],
\end{align*}
and the negativity for the minimal scalar field is
\begin{equation}
  \mathcal{N}=\frac{g_0^2}{2\pi}e^{-\Omega^2\sigma^2}(H\sigma)^2
  \left(\ln\left[\frac{2\sin(H\Omega\sigma^2)}{Hr_p}\right]
    -\frac{1}{4\sin^2(H\Omega\sigma^2)}\right).
\end{equation}
 The separability condition
$\mathcal{N}<0$ yields 
\begin{equation}
Hr_p \gtrsim
2\sin^2(H\Omega\sigma^2)\exp\left(-\frac{1}{4\sin^2H\Omega\sigma^2}\right)\sim
 1.0,
\end{equation}
where the numerical value is obtained for $H\Omega\sigma^2=1$.
Comparing this to the Minkwski vacuum case, we can find sufficiently large
$r_p$ at which detectors are separable for any value of a detector's
parameters $\Omega,\sigma$ (see Fig.~4). As $r_p$ grows in time $t_0$,
two entangled detectors $r_p< H^{-1}$ evolve to a separable state after
their separation exceeds the Hubble horizon scale. This behavior  is consistent
with our previous analysis of entanglement using the lattice model and the
coarse-grained model of the scalar field~\cite{NambuY:PRD78:2008,
  NambuY:PRD80:2009}: bipartite entanglement of the scalar field
between spatially separated regions in de Sitter spacetime disappears
beyond the Hubble horizon scale.  For the conformal invariant scalar
field, the separability condition becomes
\begin{equation}
  \label{eq:conf-scale}
 Hr_p\gtrsim 2\sin
H\Omega\sigma^2\sim 1.4.
\end{equation}
The numerical value is for $H\Omega\sigma^2=1$. Contrary to the
minimal scalar field, the entangled region extends to the super
horizon scale. Although the size of the entangled region is larger than
that of the minimal scalar field, the two detectors are separable when
their separation reaches scale \eqref{eq:conf-scale} and this behavior
is the same as the minimal scalar field.

For the entangled state, we can check the violation of the Bell-CHSH
inequality. The condition of violating the Bell-CHSH inequality after
operation of local filtering is
\begin{alignat}{2}
  &Hr_p\lesssim
  2\sin^2(H\Omega\sigma^2)\exp\left(-\frac{1}{\sin^2H\Omega\sigma^2}\right)\sim
  0.34 &\quad & \text{(minimal scalar)}, \\
  &Hr_p\lesssim \sin(H\Omega\sigma^2)\sim 0.84 &\quad &
  \text{(conformal scalar)}. \notag
  \notag
\end{alignat}
Thus, for both scalar fields, the appearance of hidden nonlocality is possible
only on  the subhorizon scale and violation of the Bell-CHSH inequality
cannot be detected on  the super horizon scale.

\begin{figure}[H]
  \centering
  \includegraphics[width=0.45\linewidth,clip]{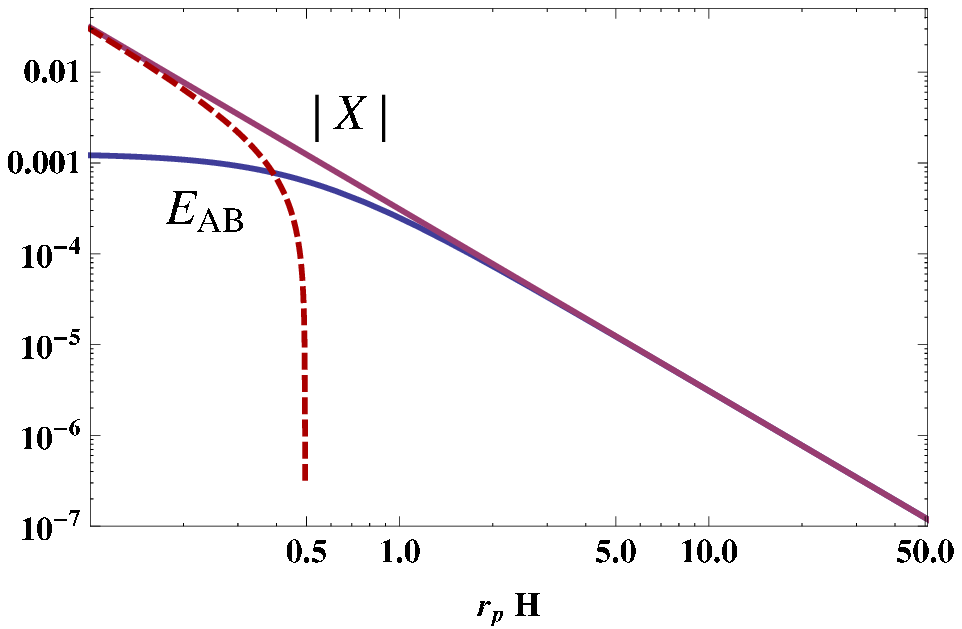}
  \includegraphics[width=0.45\linewidth,clip]{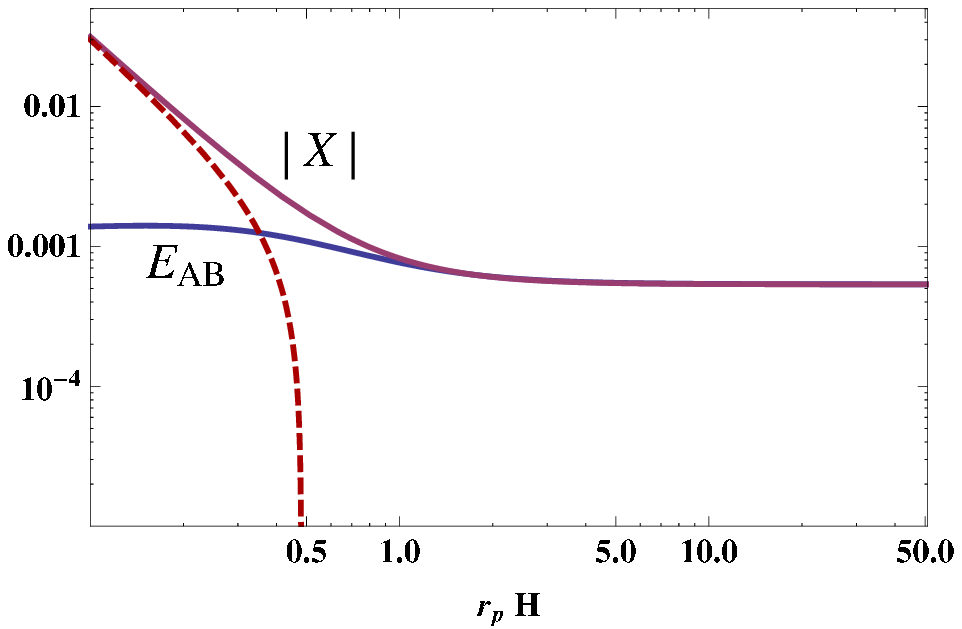}
  \caption{$r_p$ dependence of $|X|$ and $E_{AB}$ (left panel:
    conformal invariant scalar field,  right panel:  minimal scalar
    field, $H\sigma=0.5,
    \Omega/H=1, r_0=0.1, g_0=0.1$). $r_p$ is related to $t_0$ as
    $r_p=r_0e^{Ht_0}$. The dashed line represents the negativity.}
  \label{fig:EX-deS}
\end{figure}

Behavior of correlations is completely different for these
two types of scalar fields. Figure~\ref{fig:EX-deS} shows $r_p$ dependence of functions
$|X|, E_{AB}$ for the conformal invariant scalar field and the minimal
scalar field.
\noindent
For the conformal invariant scalar field, these functions decay as
$\sim r_p^{-2}$ on super horizon scale, which is the same behavior as
the Minkowski vacuum case. On the other hand, for the minimal scalar
field, these functions approach a constant value on the super horizon scale
due to accumulation of long wavelength modes of quantum
fluctuations. The different behavior of $X, E_{AB}$ leads to different
behavior of classical and quantum parts of correlations of these scalar
fields.
\begin{figure}[H]
  \centering
  \includegraphics[width=0.45\linewidth,clip]{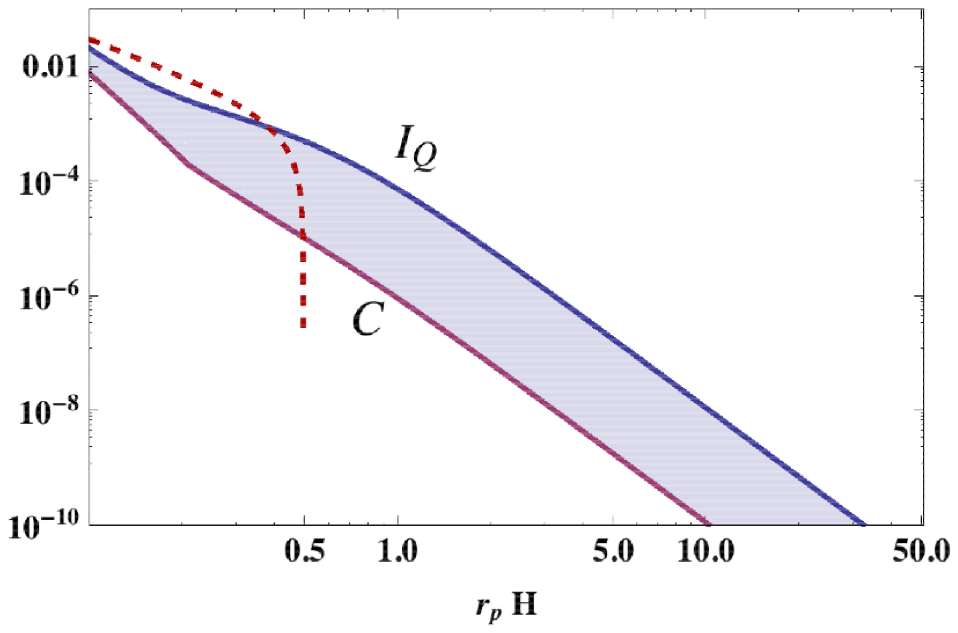}
  \includegraphics[width=0.45\linewidth,clip]{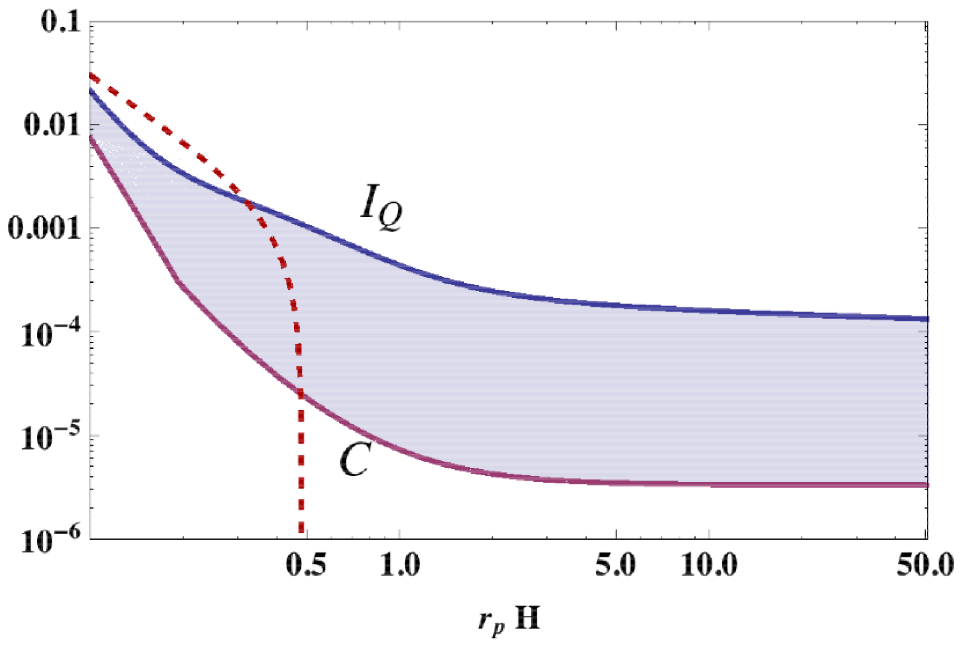}
  \caption{$r_p$ dependence of $I_{Q}$ and $\mathcal{C}$ (left panel:
    conformal invariant scalar field, right panel: minimal scalar
    field, $H\sigma=0.5, \Omega/H=1, r_0=0.1, g_0=0.1$). The dashed
    line represents the negativity.}
  \label{fig:cor-deS}
\end{figure}
\noindent
In Fig.~\ref{fig:cor-deS}, the classical and total correlation
$\mathcal{C}, I_Q$ of the minimal scalar field also approach constant
values on the super horizon scale while these correlations of the
conformal invariant scalar field decay as $\sim r_p^{-4}$.
\begin{figure}[H]
  \centering
  \includegraphics[width=0.45\linewidth,clip]{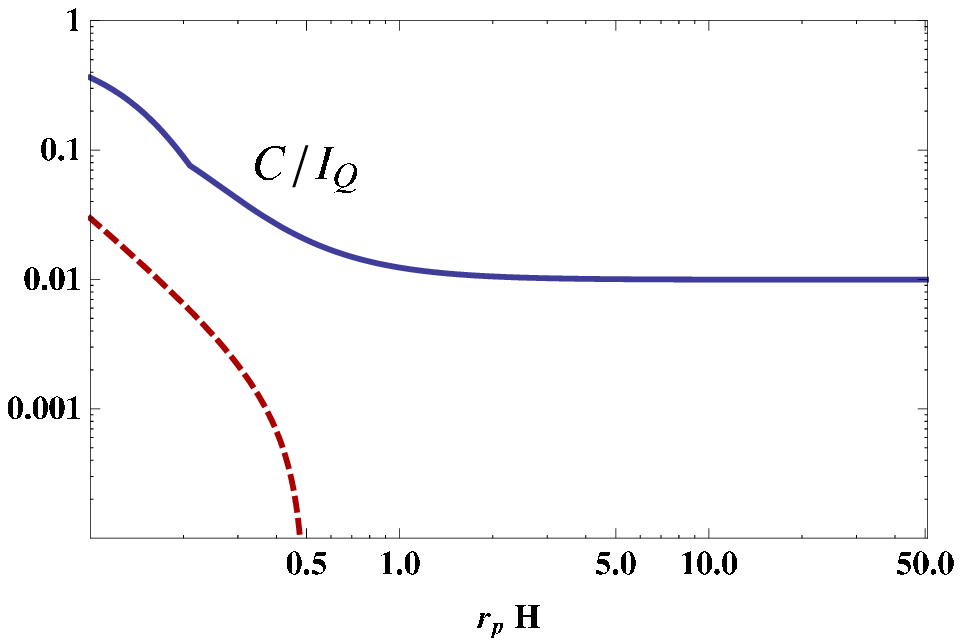}
  \includegraphics[width=0.45\linewidth,clip]{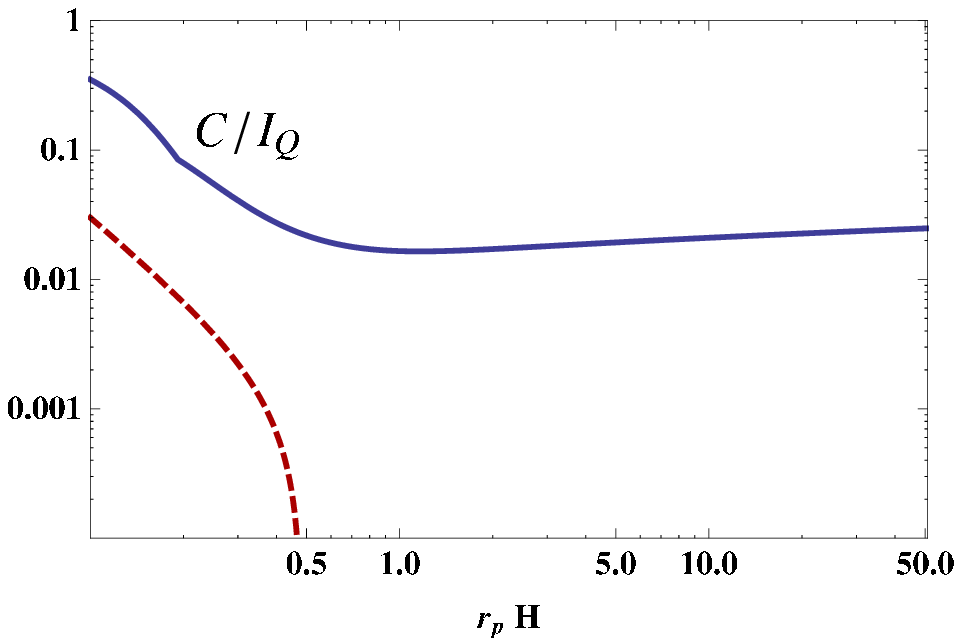}
  \caption{$r_p$ dependence of the ratio $\mathcal{C}/I_{Q}$ (left
    panel: conformal invariant scalar field, right panel: minimal
    scalar field, $H\sigma=0.5, \Omega/H=1, r_0=0.1, g_0=0.1$). The
    dashed line represents the negativity.}
  \label{fig:ratio-deS}
\end{figure}

In Fig.~\ref{fig:ratio-deS}, the ratio $\mathcal{C}/I_Q$ for the
conformal invariant scalar field approaches constant for $r_pH\gtrsim
1$ and this behavior is the same as the Minkowski vacuum case (see
Fig.~3). However, for the minimal scalar field, value of this ratio
has $\ln r_p$ dependence and increases as $r_p$ increases:
$$
 \frac{\mathcal{C}}{I_Q}\approx 4E\sim g_0^2
 e^{-(\Omega\sigma)^2}(H\sigma)^2(Ht_0)
=g_0^2 e^{-(\Omega\sigma)^2}(H\sigma)^2\ln\left(\frac{r_p}{r_0}\right).
$$
For a sufficiently large distance (large e-folding $Ht_0$) given by
\begin{equation}
  \label{eq:classical-time}
 \ln\left(\frac{r_p}{r_0}\right)\sim
 \frac{e^{(\Omega\sigma)^2}}{g_0^2(H\sigma)^2}\gtrsim 
 \frac{1}{g_0^2}\left(\frac{\Omega}{H}\right)^2,
\end{equation}
$\mathcal{C}\approx I_Q$ and this means the quantum state of detectors
approaches the zero quantum discord state and can be regarded as
 classical in the sense that measurement procedure does not alter
the state. As discussed in Ref.~\cite{ClicheM:PRA81:2010},
classical and quantum channel capacities of communication via
detectors beyond the light cone identically vanishes and we presumably
expect these capacities to be zero beyond the Hubble horizon. Thus the
correlation with zero discord on the super horizon scale originates
from quantum fluctuations of the scalar field.  Therefore, this
behavior of correlations supports the long wavelength quantum
fluctuations of the massless minimal scalar field in de Sitter
spacetime being treated as classical fluctuations.

\section{Summary}

We investigated quantum and classical correlations of the quantum
field in de Sitter spacetime using the detector model. Entanglement of
the scalar field is swapped to that of two detectors interacting with
the scalar field and we can measure the entanglement of the quantum
field by this setup of experiment.  In de Sitter spacetime, the
entanglement between detectors disappears on the super horizon scale and
this behavior is consistent with our previous analysis using the
lattice model and the coarse-grained model of the scalar
field~\cite{NambuY:PRD78:2008,NambuY:PRD80:2009}. However, the
behavior of correlations shows different behavior depending on the type of
scalar fields.  For the massless minimal scalar field, the ratio of
classical correlation to the total correlation approaches unity for
sufficiently large e-foldings. On the other hand, for the massless
conformal scalar field, that ratio approaches a constant value smaller
than unity and the condition for classicality is not achieved. These
results support the long wavelength quantum fluctuation of the minimal
scalar field being treated as classical fluctuations and  becoming
seed fluctuations for the structure in the our Universe.

As an application of our analysis presented in this paper, it is
interesting to consider quantum effects in analogue curved spacetimes
proposed using Bose-Einstein condensates or ion
traps~\cite{BarceloC:0505065:2005}. In these experiential setups of
analogue models, we can directly measure entanglement and classical
and quantum correlations of quantum fluctuation using detectors in the
laboratory. We expect that investigation in this direction will increase
understanding of the quantum and classical nature generated during the
inflation.

\begin{acknowledgments}
This work was supported in part by the JSPS Grant-In-Aid for
Scientific Research (C) (23540297).
\end{acknowledgments}

\appendix

\section{ Wightman function of massless minimal scalar field}
 In de Sitter spacetime with a spatially
flat slice, the massless minimal scalar field obeys the following equation of motion
\begin{equation}
  \ddot\phi+3H\dot\phi-e^{-2Ht}\nabla^2\phi=0.
\end{equation}
The quantized field with the Bunch-Davis vacuum state is
\begin{equation}
  \phi(t,\bs{x})=\int\frac{d^3k}{(2\pi)^{3/2}}\left(f_k(t)\,\hat
    a_{\bs{k}}+f_k^{*}(t)\,\hat
    a_{-\bs{k}}{}^{\dag}\right)e^{i\bs{k}\cdot\bs{x}}
,\quad f_k=\frac{-H}{\sqrt{2k}}\left(\eta-\frac{i}{k}\right)
 e^{-ik\eta}
\end{equation}
where $\eta=-e^{-Ht}/H$ is conformal time. The Wightman function is
\begin{align}
  D^+(x_1,x_2)&=\langle\phi(x_1)\phi(x_2)\rangle \notag \\
  &=\frac{1}{2\pi^2}\int_0^\infty
  dkk^2j_0(kr)f_k(\eta_1)f_k^*(\eta_2),\quad r=|\bs{x}_1-\bs{x}_2|
  \notag \\
  &=\frac{H^2\eta_1\eta_2}{4\pi^2r}\int_0^\infty e^{-\ep k}\sin kr
  e^{-ik\Delta\eta}
  +\frac{H^2}{4\pi^2r}\int_0^\infty dk e^{-\ep k}\sin
  kr\left(-\frac{\pa}{\pa
      k}\right)\left(\frac{e^{-ik\Delta\eta}}{k}\right)\\
&\qquad\qquad ;\Delta\eta=\eta_1-\eta_2,\notag
\end{align}
where we have introduced a damping factor $e^{-\ep k}$ with a small
positive number $\ep$ to regularize ultraviolet divergence of the 
$k$ integral.  The first integral is
\begin{equation}
  D_{\text{conf}}^{+}=\frac{H^2}{4\pi^2}\frac{\eta_1\eta_2}{-(\Delta\eta-i\ep)^2+r^2}
=\frac{H^2}{4\pi^2\bar y},\quad \bar y=\frac{-(\Delta\eta)^2+r^2}{\eta_1\eta_2}.
\end{equation}
This term is the same as the Wightman function for the massless conformal
invariant scalar field. The second integral diverges at $k=0$, hence
we introduce a lower bound $k_0$ of $k$ integral:
\begin{equation}
  D_2=\frac{H^2}{8\pi^2}\Bigl[-\mathrm{Ei}\left(-k_0(ir-i\Delta\eta-\ep)\right)
  -\mathrm{Ei}\left(-k_0(-ir-i\Delta\eta-\ep)\right)\Bigr].
\end{equation}
As the value of $k_0$, we choose $k_0=H$ which corresponds to the size
of the universe (horizon scale) at the beginning of inflation.

\end{document}